# Evaluating Earth-Observing Satellite Sampling Effectiveness Using Kullback-Leibler Divergence

Negin Esmaeili[a], Paul T. Grogan[a]

[a] School of Computing and Augmented Intelligence Arizona State University, Tempe, AZ, USA, nesmaei2@asu.edu ,Paul.Grogan@asu.edu

*Abstract*

*This work presents an objective, repeatable, automatic, and fast methodology for assessing the representativeness of geophysical variables sampled by Earth-observing satellites. The primary goal is to identify and mitigate potential sampling biases attributed to orbit selection during pre-Phase A mission studies. This methodology supports current incubation activities for a future Planetary Boundary Layer observing system by incorporating a sampling effectiveness measure into a broader architectural study. The study evaluates the effectiveness of 20 satellite configurations for observing convective storm activity in the Southwestern U.S. during the North American Monsoon (NAM) season. The primary design variables are the number of satellites, orbit type (sun-synchronous or inclined), and Local Time of Ascending Node (LTAN). Using Kullback-Leibler (KL) divergence to assess observational representativeness and Kernel Density Estimation (KDE) to estimate probability density functions, the study quantifies the discrepancy between observed and ground truth storm features. Results indicate that a two-satellite sun-synchronous system with an 8:00 PM LTAN, achieved the lowest KL divergence, signifying the most representative observation of storm clusters. In contrast, single-satellite configurations, particularly those with late-night LTANs (e.g., 12:00 AM), demonstrated significantly higher KL divergence. The study concludes that dual-satellite configurations in sun-synchronous orbits with evening LTANs outperform single-satellite and inclined configurations in capturing representative convective storm activity.*

## 1. Introduction

The Planetary Boundary Layer (PBL) is the lowest turbulent layer of the atmosphere, directly influenced by its contact with the Earth's surface. The PBL varies geographically and temporally, spanning diurnal to seasonal scales, and influences the global weather and climate system. Observing the PBL is vital for understanding pollutant dispersion and climate modeling, as it governs weather systems like convective storms, which can produce extreme weather events such as thunderstorms, squall lines, and tropical cyclones [1]. Convective storms drive atmospheric dynamics through the vertical transport of moisture and energy in heavy rain, thunderstorms, and tornadoes. Storms are influenced by the PBL's thermodynamic structure, affecting their onset, intensity, and duration. They also affect precipitation distribution, atmospheric constituent mixing, air quality, and human health [2]. Understanding weather processes is crucial for forecasting, air quality assessments, and climate research. Furthermore, the diurnal cycle of convection highlights PBL's dynamic role in storm initiation and evolution [3].

This paper is part of a broader effort to design future satellite missions by developing an end-to-end Observing System Simulation Experiment (OSSE) framework. Our project seeks to evaluate PBL observation mission concepts, with a focus on convective storms. The primary objective is to provide decision support through a methodology that evaluates the effectiveness of observing systems by integrating satellite simulations and geophysical datasets. Traditional mission evaluation relies heavily on subjective assessments involving human interpretation combined with computational tools. In contrast, our approach makes it more objective, automatic, and fast, ensuring transparency, repeatability, and efficiency. This methodology enables rapid and thorough assessment of the representativeness of satellite observations in capturing key PBL features, such as convective storm activity, enhancing scientific value.

Our work proposes a metric that evaluates the representativeness of satellite observations with respect to ground truth geophysical phenomena, specifically convective storms within the PBL. We developed this metric to objectively measure how well satellite configurations contribute to PBL observations, providing insights to improve mission design. Our proposed methodology integrates engineering simulations of satellite orbits with geophysical datasets to assess satellite configuration's ability to observe the phenomena of interest of the PBL, comparing ground tracks with ground truth convective activity.

An application case study focuses specifically on observing convective storm activity during the North American Monsoon (NAM) in the U.S. Southwest. The NAM, a challenging region with intense storm activity, is a prime test case for evaluating the performance of satellite observing systems. Our proposed method utilizes Kullback-Leibler (KL) divergence and Kernel Density Estimation (KDE) to analyze satellite configuration effectiveness to collect representative observations. Analysis results offer valuable insights for future Earth-observing missions design and guide future decisions based on satellite configurations ability to capture atmospheric processes within the PBL.

## 2. Background

### 2.1. Trade Studies

Trade studies are essential in satellite mission design, offering a systematic approach to evaluate and compare design alternatives. These studies optimize mission performance, manage resources, and mitigate risks by balancing scientific objectives, cost, schedule, and technical feasibility. They provide a structured framework for decision-making, enabling mission planners to assess various configurations and strategies, resulting in more effective Earth-observing satellite missions [4]. Several large-scale trade studies exemplify this systematic approach. The Aerosols, Clouds, Convection, and Precipitation (ACCP) study [5], the NOAA Satellite Observing System Architecture (NSOSA) trade study [6], and the Landsat Mission Analysis [7] are key examples of comprehensive satellite mission evaluations.

The ACCP study narrowed over 100 potential satellite architectures to the top three by focusing on variables such as assessing geophysical measurement quality and cost and recommending systems for observing atmospheric phenomena. The ACCP study underscores the importance of evaluating numerous architectures, assessing trade-offs, and recommending optimal configurations to guide mission design.

The NOAA Satellite Observing System Architecture (NSOSA) trade study explored more than 180 satellite constellations, identifying the most cost-effective configurations that would meet the needs of numerical weather prediction (NWP) for environmental monitoring, using OSSEs to evaluate their impact. Key outcomes included recommendations for optimizing satellite orbits, increasing the number of radio occultation (RO) soundings, and enhancing the revisit rate for satellite radiance soundings.

The Landsat Mission Analysis represents another trade study to ensure the continuity and enhancement of the Landsat program's Earth observation capabilities by evaluating satellite architectures, sensor technologies, and operational approaches for enhancing Earth observation capabilities. It aimed to select a satellite configuration that balanced scientific objectives and budget, exemplifying trade-offs in mission design for applications in agriculture, climate monitoring, and land use.

### 2.2. Decision Analysis Tools

Decision analysis tools like The Advanced Systems Performance Evaluation Tool for NOAA (ASPEN) and Value Assessment Simulation for Space Architecture Redesign (VASSAR) provide structured technical approaches to integrate multi-attribute data within trade studies. ASPEN, designed to optimize Earth-observing systems, evaluates how well alternative observing systems meet user-defined requirements by analyzing trade-offs between designs, estimating cost- effectiveness, and conducting "what-if" simulations to test system modifications to support decision-making across domains like meteorology and oceanography [8]. VASSAR helps mission planners assess the relative merit of space mission architectures by simulating designs and evaluating performance against quantitative and qualitative factors like coverage, data quality, and cost, enabling a balanced approach to performance and stakeholder needs [9].

### 2.3. Mission Evaluation Tools

OSSEs assist engineers and scientists in designing, analyzing, and validating satellite missions. They simulate satellite observations by integrating geophysical datasets with operational processes and help predict observational impacts on weather forecasts and climate models [10]. OSSEs evaluate trade-offs in satellite mission design, helping quantify potential benefits before launch, enabling design optimization to improve weather analysis and prediction. The Tradespace Analysis Tool for Constellations (TAT-C) helps mission planners evaluate satellite configurations by simulating performance metrics like coverage, revisit time and calibration accuracy. TAT-C integrates with high-resolution datasets like the GEOS-5 Nature Run (G5NR), which provides global, detailed weather simulations, including convective storms, allowing comprehensive assessments of constellation performance to support mission design decisions [11, 12].

2.4. Research Gap and Objective

The research gap in current methodologies stems from the lack of tools capable of quantitatively assessing the sampling effectiveness of Earth-observing satellite systems, particularly in capturing geophysical phenomena such as convective storm events within the PBL. Mission evaluation tools help evaluate satellite orbits, coverage, and mission performance, but lack the capability to statistically quantify how well satellite observations represent underlying geophysical phenomena and capture true geophysical features. Traditional trade studies often rely on expert judgment and qualitative assessments, which can introduce biases in evaluating mission effectiveness. These methods lack a consistent way to quantitatively assess how well satellite observations represent true geophysical phenomena. A quantitative approach, like KL divergence, provides an objective, repeatable metric for assessing observation accuracy, reducing subjectivity, and improving mission design transparency.

The objective of this research is to develop a quantitative, automated methodology for assessing the ability of Earth-observing satellite systems to observe representative geophysical phenomena, such as convective storms within the PBL. This approach quantitatively compares the effectiveness of different satellite configurations to mitigate sampling biases from orbit selection. By identifying optimal configurations that minimize observational discrepancies, this framework provides mission planners with a tool to enhance the design and performance of future satellite missions.

**3. Methodology**

Our methodology, illustrated in Figure 1, defines several steps to create a robust framework for evaluating observing system representativeness. It from nature run data to event-based representations by identifying and clustering key features, such as storm events. Different satellite configurations are then evaluated for their ability to observe these features, using KDE and KL divergence to quantify representativeness against the true data distribution.

3.1. Nature Run Dataset

The study uses the GEOS-5 Nature Run (G5NR) dataset managed by NASA's Global Modeling and Assimilation Office (GMAO) [13,14], as a source of high-resolution simulated ground truth geophysical variables. This dataset includes spatial (latitude and longitude) and temporal (time) dimensions, defined on a global grid covering the geographic region. The spatial grid is defined by longitude indices $i \in \{1, ..., I\}$ and latitude indices $j \in \{1, ..., J\}$. We denote the feature value at grid cell $(i, j)$ at time $t$ as $G_{i,j}(t)$ which corresponds to a specific location and captures the geophysical variables at a given time $t$. For this application, we focus on the North American Monsoon (NAM) season in the U.S. Southwest, specifically targeting the region from California to Texas (covering parts of CA, NV, UT, AZ, CO, NM, and TX). The NAM is known for its intense, short-lived convective storms that significantly impact the region by contributing to seasonal rainfall and occasionally causing extreme weather events. The U.S. Southwest is an ideal location for studying these phenomena due to the high frequency of such storms during the monsoon season. The rapid development and dissipation of these storms, coupled with their spatial variability, present significant challenges for Earth-observing (EO) satellite systems. Capturing critical moments in their evolution requires precise and timely satellite coverage. This variability makes the study of these storm systems valuable for tradespace analysis, emphasizing the trade-offs between different satellite configurations in effectively capturing rapidly evolving weather phenomena. The study covers a two-month period from July 15th to September 15th, which coincides with the peak of the monsoon season.

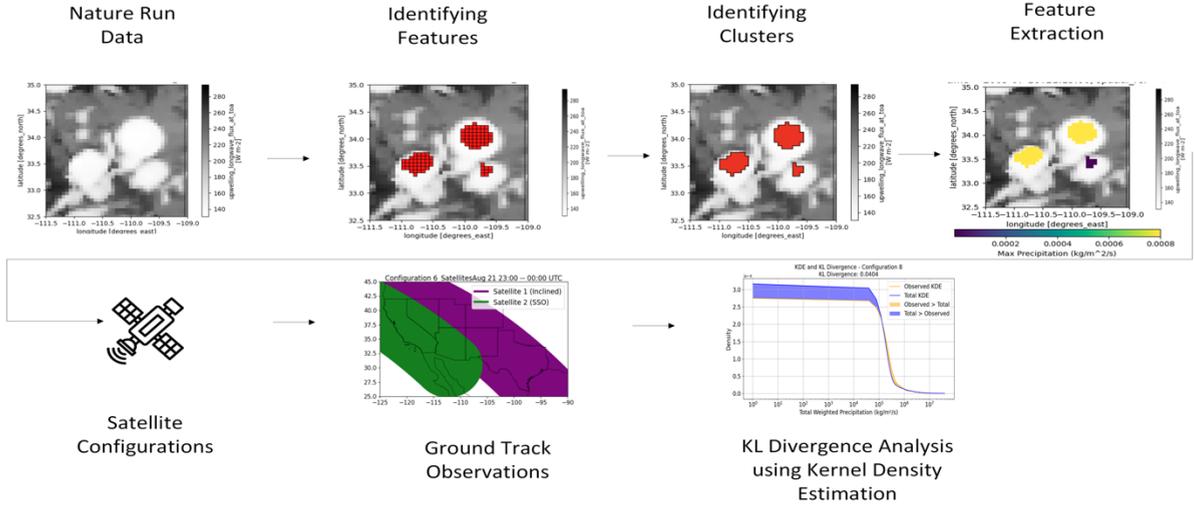

Fig. 1. Methodological steps to evaluate the effectiveness of observing systems.

The primary geophysical variable is total precipitation (PRECTOT), which helps in identifying and characterizing convective storm events in the dataset. We also use longwave radiation at the top of the atmosphere (LWTUP) to detect cloud-top temperatures, which are indicative of intense convective activity.

3.2. Feature Identification and Clustering

The first step involves identifying storm features from the gridded geophysical data provided by the G5NR dataset, which includes geophysical variables such as total precipitation and longwave radiation. A storm feature is identified based on its intensity and spatial extent. For this purpose, we focused on cold cloud top temperatures, indicators of intense convective activity, which can be identified using LWTUP values. A threshold cloud top temperature of 220 K is equivalent to $5.67037 \cdot 10^{-8} \cdot 220^4 = 132.8$ W/m² via the Stefan-Boltzmann law. A binary mask isolates grid cells where LWTUP was below the threshold value. The cold cloud mask is a binary array where cells with cloud top temperatures below 220 K are labeled as 1, and all other cells are labeled as 0.

The next step applies a clustering algorithm, which groups adjacent grid cells that share similar geophysical properties. The binary mask was passed through a connected-component labeling algorithm using the label function from the scipy.ndimage module. This function scans the binary array and forms clusters of adjacent cells that meet the threshold requirement, assigning a unique label to each cluster. As illustrated in Figure 1, the labeling process is based on 4-connectivity (edge adjacency) in a 2D grid. Mathematically, the labeling process defines the cluster index $C_k(t)$, where $k$ is the unique label assigned to a connected component, and $t$ represents the time step.

Here, $C_k$ represents the $k$-th cluster identified by the scipy.ndimage.label function. This method allows us to systematically identify and label contiguous regions of significant precipitation, which are then used for further analysis and attribute calculation. To represent the geospatial extent of each cluster, we converted the grid cells within each cluster into polygons using the geopandas library. Each grid cell was represented as a polygon, and the entire cluster was aggregated into a multi-polygon representing its spatial boundary. The centroid of each cluster was also calculated to provide a spatial reference for the storm event. Finally, the identified clusters, along with their spatial properties and precipitation attributes, were stored in a GeoDataFrame. As shown in Figure 1, this transformation of raw data into event-based clusters provided a detailed view of convective storm features, serving as the ground truth for satellite observation analysis and design optimization. This process allows us to identify, track, and analyze clusters over time, attributing geophysical data to discrete features, such as storm events, for further analysis.

3.3. Feature Extraction

After clustering, the feature extraction process calculates attributes for each identified cluster based on the constituent nature run geophysical variable values. These extracted features provide information about the intensity and spatial extent of each event. Cluster $C_k(t)$ denotes the $k$-th cluster at time $t$, consisting of multiple grid cells $(i,j) \in C_k(t)$. For each grid cell within the cluster, we retrieve the feature value $G_{i,j}(t)$ at time $t$, which represents the geophysical variable of interest $G$ (e.g., precipitation or temperature). Based on the feature values retrieved for all grid cells within the cluster, we compute cluster attributes $G_{x,k}(t)$ to characterize attribute $x$ for features of interest. Here, $x$ refers to an aggregation function such as maximum value, total value, or average value of the cluster.

Maximum Value- $G_{\max,k}(t)$ identifies the peak intensity of $G$ within cluster $k$ at time $t$.

$$G_{\max,k}(t) = \max_{(i,j) \in C_k(t)} G_{i,j}(t) \quad (1)$$

Total Weighted Value- $G_{\text{total},k}(t)$ sums the feature values $G_{i,j}(t)$ at time $t$ over all grid cells $(i,j)$ in the cluster, weighted by the area $A_{i,j}$ of each grid cell calculated based on the geographic coordinates of the grid cell.

$$G_{\text{total},k}(t) = \sum_{(i,j) \in C_k(t)} G_{i,j}(t) \cdot A_{i,j} \quad (2)$$

Average Value- $G_{\text{avg},k}(t)$ provides a measure of the overall intensity across the cluster. This is useful for comparing the feature's intensity across clusters of varying sizes, helping to assess whether the feature's impact is concentrated or more evenly distributed. It is calculated by:

$$G_{\text{avg},k}(t) = \frac{1}{|C_k(t)|} \sum_{(i,j) \in C_k(t)} G_{i,j}(t) \quad (3)$$

where $|C_k(t)|$ is the number of grid cells in cluster $C_k(t)$.

Feature extraction calculates attributes for each cluster identified during the clustering step. These attributes characterize the storm features in terms of their intensity, spatial extent, and overall contribution to the total precipitation in the region. The main attribute computed is the total precipitation within each cluster. This attribute provides an understanding of each storm event's intensity and spatial distribution. We adapted Equation (2) to calculate the total weighted precipitation for each identified cluster $C_k(t)$ by summing the precipitation values over all grid cells within the cluster. Each grid cell's contribution is weighted by its respective area, ensuring that larger cells contribute proportionally to their size. This approach is crucial for assessing the overall contribution of the cluster to the total precipitation in the region. In this way, this extracted attribute — total precipitation—reflects the storm's overall contribution to regional rainfall. This feature value was stored for each cluster in the dataset, allowing for further analysis.

3.4. Satellite Configuration and Observability: Experimental Design

After feature extraction, we evaluated the ability of various satellite configurations to observe the identified storm clusters. The goal was to assess the effectiveness of different orbits in capturing convective storm events and to quantify potential sampling biases, allowing us to identify the most effective setup for storm observation. For this study, we evaluated 20 satellite configurations, varying across two key variables: the number of satellites and the orbit type. The number of satellites was either one or two, and the orbit type included both inclined orbits (50° or 55° inclination) and sun-synchronous orbits with local time of ascending node (LTAN) values of 8:00 PM, 10:00 PM, or 12:00 AM. For two-satellite configurations, if both satellites were placed in the same orbit, the true anomaly of the second satellite was set to 180 degrees. This ensured that the satellites were positioned on opposite sides of the orbit, maximizing spatial separation and improving overall coverage. To maintain consistency, the altitude was fixed at 700 km for all configurations, and each satellite was equipped with the same instrument featuring a swath width of 1,450 km. By holding both the altitude and instrument parameters constant, we isolated the impact of the orbital variables—specifically the number of satellites, inclination, and LTAN—allowing us to directly attribute performance differences to these factors.

3.5. Observability Evaluation Process

To evaluate the observability of the storm clusters, satellite ground tracks were computed for each configuration over the study period, from July 15, 2005, to September 15, 2005, using the TAT-C library. The ground tracks were computed at regular intervals (every 30 minutes), providing a high temporal resolution to match the frequency of the storm clusters. After generating the satellite ground tracks, analysis determined whether the satellites observed each storm cluster. This process involved both temporal alignment and spatial matching of the satellite ground tracks with the storm clusters. First, the timestamps of the storm clusters were compared with the timestamps of the satellite ground tracks. If a satellite pass occurred at the same time as the storm cluster, the next step was to check the spatial alignment. We evaluated whether the centroid of each storm cluster fell within the satellite's field of view during the pass. If the centroid was within the satellite's field of view at the correct time, the cluster was marked as observed. The output of this experiment was a dataset indicating whether each storm cluster was observed by the satellites in the various configurations. For each configuration, the clusters were classified as either observed or not observed, based on the spatial and temporal matching with the satellite ground tracks.

3.6. Kernel Density Estimation

After identifying the storm clusters and conducting the observability experiment, we applied KDE to estimate the probability density function (PDF) of total weighted precipitation for both the complete set of clusters and the subset observed by each satellite configuration. KDE, also known as Parzen's window [15], is a commonly used non-parametric method for estimating the distribution of a dataset, making no assumptions about its underlying distribution [16]. KDE effectively handles both discrete and continuous data by smoothing observations to produce an estimated distribution. For continuous variables, KDE approximates the PDF commonly using Gaussian kernel, which gives more weight to nearby data points and gradually reduces the influence of distant ones, creating a smooth, continuous estimate of the distribution. This flexibility and computational efficiency make KDE widely applicable across various data types. By adjusting weights for data points based on their distance from the target, KDE achieves accurate density estimation even when the true distribution is unknown. While KDE can work with different kernel functions, the choice of kernel has minimal impact on the estimation accuracy. While KDE can use various kernel functions with minimal impact on estimation accuracy, the choice of bandwidth h is more important as it significantly affects smoothing. We used Silverman's reference bandwidth, which balances smoothing and is robust to outliers, ensuring reliable density estimates even with extreme values [17].

In KDE, boundary bias is a common issue, especially when the data distribution has a natural boundary [18]. Some geophysical data, such as precipitation in our case cannot be negative. This bias occurs because the kernel function extends beyond the feasible range of the data, leading to an underestimation of the density near the boundary. For this study, we addressed this boundary bias by employing a mirroring technique. The mirroring technique works by reflecting data around the boundary. For each precipitation data point, a mirrored counterpart is created. This approach effectively doubles the dataset, creating a symmetric distribution around zero, allowing KDE to better capture density near the boundary. Figure 2 illustrates the mirroring process: the KDE for the original data (red curve) drops off near zero, while the mirrored histogram (blue) reflects data around the boundary, correcting the bias. The combined KDE from both original and mirrored data (orange) provides an improved boundary estimate, while the final KDE (blue dashed curve) offers a smooth, bias-corrected density representation.

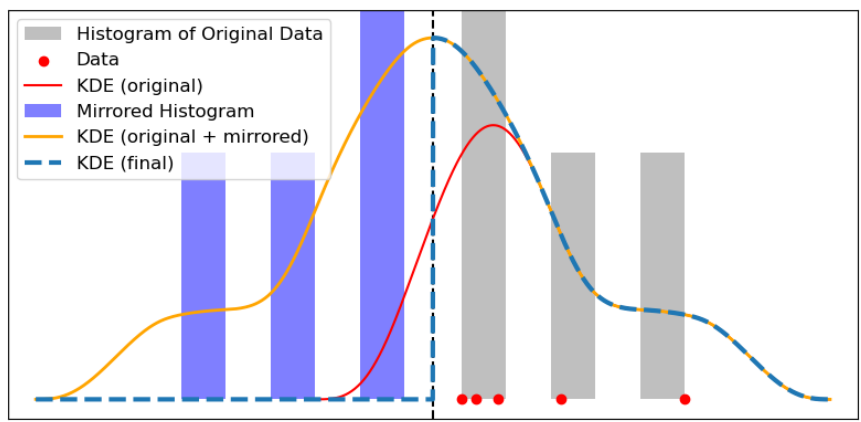

Fig. 2. Conceptual illustration of the mirroring technique

### 3.7. Kullback-Leibler (KL) divergence

To evaluate how well satellite configurations observe convective storm activity, we applied Kullback-Leibler (KL) divergence. The KL divergence was first introduced in 1951 [18]. KL divergence is a measure of how one probability distribution diverges from a second reference probability distribution. This comparison assesses the difference between the observed storm clusters (captured by the satellites) and the total storm clusters (representing the full dataset). For continuous distributions f(x) and g(x), the KL divergence is defined as (Equation 4):

$$D_{\text{KL}}(f\|g) = \int_{-\infty}^{\infty} f(x)\log \frac{f(x)}{g(x)} dx \quad (4)$$

where $f(x)$ represents the distribution of the observed storm features by the satellite constellation and $g(x)$ represents the ground truth distribution of convective storm features. KL divergence is asymmetric and measures the information lost when $g(x)$ is used to approximate $f(x)$. A $D_{\text{KL}}(f\|g) = 0$ indicates that the observed distribution $f(x)$ perfectly matches $g(x)$, implying that the satellite observations are fully representative of the actual storm features. Conversely, $D_{\text{KL}}(f\|g) > 0$ indicates a positive KL divergence, meaning a discrepancy between the observed and true distributions. Larger values signify greater divergence, suggesting that the satellite observations are less representative of the actual storm features.

To focus on the most significant storm clusters and reduce the influence of extreme outliers, we set the upper integration bound in Equation 4 to the 99th percentile of the total weighted precipitation values, denoted as $P_{99}$. The lower bound is set as the 0th percentile, denoted as $P_0$, representing the minimum observed value. Numerical integration over this range, from $P_0$ to $P_{99}$, allowed us to compute the KL divergence, providing a quantitative score for each satellite configuration.

## 4. Results

Analysis enabled us to rank satellite configurations by their KL divergence values, with lower values indicating configurations that more representatively observe full storm activity. This ranking provides a metric for comparing observational effectiveness of different configurations in representing storm features. Table 1 summarizes the KL divergence results, ranking the satellite configurations accordingly.

Table 1. Ranking of Satellite Configurations Based on KL Divergence. "x" indicates the number of satellites (e.g., 1x for one satellite). SSO refers to Sun-Synchronous Orbit, and LTAN represents the Local Time of Ascending Node.

| Rank | Configuration | Details | KL Divergence | Rank | Configuration | Details | KL Divergence |
|---|---|---|---|---|---|---|---|
| 1 | 8 | 2x SSO 8:00 PM LTAN | 0.0404 | 11 | 4 | 1x SSO 10:00 PM LTAN | 0.0912 |
| 2 | 18 | 1x SSO 8:00 PM LTAN, 1x SSO 10:00 PM LTAN | 0.0494 | 12 | 20 | 1x SSO 10:00 PM LTAN, 1x SSO 12:00 AMLTAN | 0.0928 |
| 3 | 9 | 2x SSO 10:00 PM LTAN | 0.0604 | 13 | 14 | 1x 50° Inclination, 1x SSO 12:00 AMLTAN | 0.1319 |
| 4 | 12 | 1x 50° Inclination, 1x SSO 8:00 PM LTAN | 0.0644 | 14 | 1 | 1x 50° Inclination | 0.1344 |
| 5 | 13 | 1x 50° Inclination, 1x SSO 8:00 PM LTAN | 0.0683 | 15 | 7 | 2x 55° Inclination | 0.1744 |
| 6 | 15 | 1x 55° Inclination, 1x SSO 8:00 PM LTAN | 0.0729 | 16 | 11 | 1x 50° Inclination, 1x IO55° Inclination | 0.1584 |
| 7 | 3 | 1x SSO 8:00 PM LTAN | 0.0772 | 17 | 10 | 2x SSO 12:00 AM LTAN | 0.1643 |
| 8 | 19 | 1x SSO 8:00 PM LTAN, 1x SSO 12:00 AMLTAN | 0.0811 | 18 | 17 | 1x 55° Inclination, 1x SSO 12:00 AMLTAN | 0.1688 |
| 9 | 16 | 1x 55° Inclination, 1x SSO 10:00 PM LTAN | 0.0849 | 19 | 2 | 1x 55° Inclination | 0.1820 |
| 10 | 6 | 2x 50° Inclination | 0.0911 | 20 | 5 | 1x SSO 12:00 AM LTAN | 0.2332 |

The results of the KL divergence analysis show that Configuration 8, which includes two satellites in sun-synchronous orbits with an 8:00 PM LTAN performed best among the other configurations, with a KL divergence value of 0.0404. This is shown in Figure 3 where the observed and total KDE curves closely align, indicating a strong correspondence between the observed storm clusters and the total storm activity. The small difference between the observed and total KDE curves suggests that Configuration 8 provides strong coverage of storm events during the monsoon period.

In contrast, Configuration 5, a single satellite in a sun-synchronous orbit with a 12:00 AM LTAN, had the highest KL divergence of 0.2332, as illustrated in Figure 4 The large shaded blue area between the observed and total KDE curves indicates that this configuration fails to capture a substantial portion of the storm activity, particularly afternoon and early evening convective storms, which are more intense and frequent during the monsoon period.

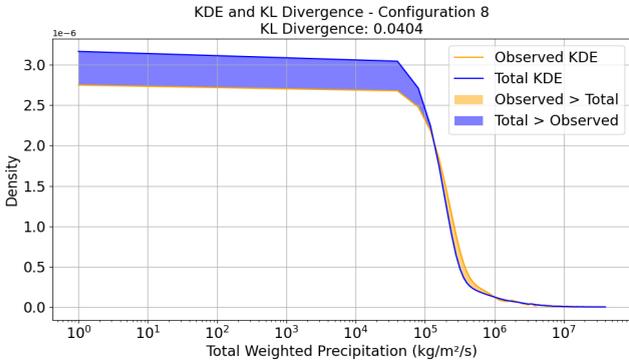
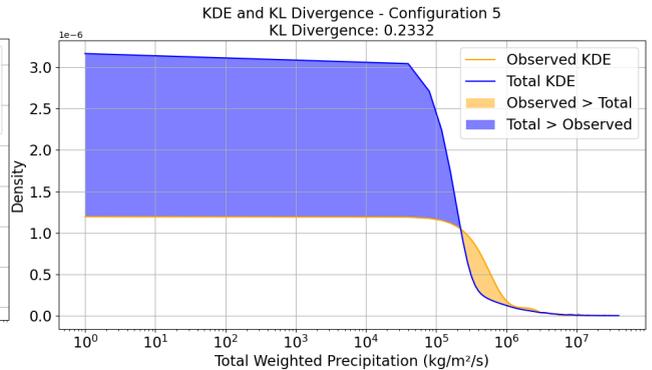

Fig. 3. Observed and Total KDE Curves for Configuration 8

Fig. 4. Observed and Total KDE Curves for Configuration 5

This divergence highlights the limited effectiveness of Configuration 5 in observing the full extent of the convective storm activity, especially during the peak times of convective storm development. These results provide a ranking of the satellite configurations, as shown in Table 1, based on their ability to observe convective storm clusters representatively. Dual-satellite setups, especially those in sun-synchronous orbits with evening LTAN, performed better in capturing storm activity compared to single-satellite setups, indicating stronger representativeness of storm clusters. Inclined orbits, while offering broader temporal coverage, generally showed higher KL divergence values. However, configurations that combine inclined orbits with sun-synchronous orbits demonstrated a balanced performance, achieving more consistent coverage than purely inclined setups. This highlights the importance of orbit selection and timing in determining the effectiveness of satellite observations during the North American Monsoon season.

## 5. Discussion

### 5.1. Key Considerations for Satellite Mission Design

Dual-satellite configurations are generally more effective at improving observational representativeness compared to single-satellites, as they provide greater spatial coverage and higher revisit rates. These setups are especially valuable for observing rapidly evolving, short-lived weather events like convective storms, which require frequent observations to capture their full life cycle. In this study, configurations utilizing two satellites, particularly in sun-synchronous orbits with evening LTANs, such as Configurations 8, 18, and 9, achieved the lowest KL divergence values, meaning they closely matched the true distribution of storm clusters. This demonstrates that dual-satellite setups provide more representative observations of convective storm clusters, especially during concentrated periods such as late afternoon and early evening.

The timing of satellite orbits plays a crucial role in maximizing observational effectiveness. Sun-synchronous orbits, which maintain consistent lighting conditions, pass over the same location on Earth at approximately the same local solar time each day. This makes them particularly advantageous for observing weather patterns that follow diurnal cycles. In this study, sun-synchronous orbits with evening LTANs, such as Configuration 8 (8:00 PM LTAN), aligned well with peak convective storm activity driven by diurnal heating patterns. This alignment ensures that the satellites capture the most significant storm events during the monsoon season.

Inclined orbits provide broader temporal coverage throughout the day, offering the advantage of observing different regions at various times. However, their performance may be limited when specific time-sensitive phenomena, like convective storms, are concentrated within narrow time windows. In this study, inclined orbits underperformed in capturing storm events during peak convective times, leading to missed observations. However, performance improved when inclined orbits were combined with sun-synchronous orbits, as seen in Configurations 12, 13, and 15, by leveraging the broad temporal coverage of inclined orbits and the optimal timing of sun-synchronous orbits.

5.2 Limitations of the Application Case

The application case in this study is constrained by several factors that limit the generalizability of the results. First, the study focuses on a two-month period from July 15 to September 15 during a single year. Storm activity varies significantly from year to year, influenced by larger climatic phenomena like El Nino and La Nina. Expanding the analysis to multiple years or the full monsoon season would provide a more robust assessment of the satellite configurations' performance.

Second, the satellite configurations in this study are limited to 20 specific setups with fixed altitudes (700 km) and identical instrument parameters. This restricted selection does not fully capture the range of possible satellite configurations. Satellites often operate at varying altitudes, with different instrument sensitivities and swath widths, all of which could influence the effectiveness of storm observation.

Lastly, while the study optimized sun-synchronous orbits for storm observations by choosing evening LTANs (e.g., 8:00 PM), most operational Earth-observing satellites use standard LTANs around 1:30 PM or 1:00 AM to meet broader observational requirements globally. Thus, while these specialized orbits are effective for storm observation, they may not be feasible for all missions.

5.3. Limitations of the Method

While the methodology offers a systematic framework for evaluating satellite observing system effectiveness, certain limitations should be considered. One key limitation is the reliance on univariate analysis using total weighted precipitation as the sole attribute for KL divergence calculations. Convective storms are influenced by multiple variables, such as temperature, humidity, and wind patterns, which were not captured in this study. Future work should incorporate multivariate analyses to capture the full complexity of storm characteristics. Another limitation is the assumption that satellite configurations are evaluated equally based on their observational representativeness. Real-world missions involve trade-offs among various factors, including cost, satellite lifespan, and instrument capabilities, which are not addressed in this analysis. Including these additional factors would improve the applicability of the findings for mission decision-making.

## 6. Conclusion

This study presents an automated, objective methodology for evaluating the effectiveness of Earth-observing satellite configurations using KL divergence. By comparing the observed and ground truth distributions of storm activity, the methodology quantifies how well different satellite configurations represent the underlying geophysical features. Results demonstrate that configurations with two satellites and sun-synchronous orbits with evening LTANs are the most effective in capturing convective storm events in the Southwest U.S. In contrast, single- satellite setups and inclined orbits perform less effectively.

Future research should broaden the analysis to include a wider array of satellite configurations, incorporating variations in altitude, instrument characteristics, and orbital parameters. Extending the temporal scope to cover multiple years and seasons would provide insights into the consistency of the configurations' performance under different climatic conditions. Incorporating a multivariate analysis framework would allow for a more holistic evaluation of storm events, capturing the interplay between various atmospheric variables. Integrating practical considerations such as cost analysis, satellite longevity, and operational feasibility into the evaluation process would enhance the methodology's applicability for real-world mission planning.

**Acknowledgment**

This work is supported by a NASA Planetary Boundary Layer Decadal Science Incubator project under subcontract 1704657 from the Jet Propulsion Laboratory.